# NEW EXPERIMENT ON THE NEUTRON RADIATIVE DECAY

<u>Khafizov R.U.</u>[a], Kolesnikov I.A[a]., Tolokonnikov S.V.[a], Torokhov V.D.[a],
Solovei V.A.[b], Kolhidashvili M.R.[b],
Konorov I.A.[c]

[a] RRC Kurchatov Institute, Moscow, Russia
[b] Petersburg Nuclear Physics Institute, Gatchina, Russia
[c] Technical University of Munich, Munich, Germany

Abstract

The report is dedicated to the preparation of the new experiment on the neutron radiative decay what is conducted for the last years. We started the experimental research of this neutron decay branch with the experiment conducted at ILL in 2002 and continued in another experiment at the second and third cycles at the FRMII reactor of the Technical University of Munich in 2005. In the first experiment we succeeded in measuring only the upper limit on the relative intensity (B.R.) of the radiative neutron decay and in the second we succeeded in discovering events of radiative neutron decay and measure its B.R.=$(3.2\pm1.6)10^{-3}$ ( with C.L.=99.7% and gamma quanta energy over 35 keV ). The obtained average B.R. value was approximately twice the theoretical value calculated earlier within the framework of the standard electroweak model. However, due to significant experimental error it would be preliminary to deduce that based on this finding a deviation from the standard model has been observed. To prove or disprove the existence of a deviation it is necessary to conduct a new experiment that would allow to measure the radiative peak in timing spectra with precision in the order of 1%. By the present time we have prepared a new experiment the main result of which would be the measurement of B.R. for the radiative branch of neutron decay with this precision.

Among the many branches of elementary decay with charged particles in the final state, the radiative branch, where the decay occurs with the creation of an additional particle – the gamma quantum, is usually the most intensive, as the relative intensity ( or branching ratio B.R. ) of this mode is determined by the fine structure constant $\alpha$ of $10^{-2}$ order of magnitude. This decay branch is well established and was investigated for almost all elementary particles. However, the radiative decay of the free neutron

$$n \rightarrow p + e + \bar{\nu} + \gamma$$

was not discovered, and all the experiments were aimed at the study of the ordinary neutron decay branch. We conducted the first experiment on the discovery of this rare neutron decay branch in 2002 on the intensive cold neutron beams at ILL (Grenoble, France), where we received a limit for the relative intensity of this rare decay branch: B.R. < $6.9*10^{-3}$ ( 90% C.L.) [8]. This value exceeds the theoretical value we calculated and published only in by a few times. This fact, in turn, means that in our experiment of 2002 we came very close to discovering the radiative mode of neutron decay. For reasons outside of our control we did not receive beam time on intensive cold neutron beam at ILL for a number of years and were able to continue our experiment only in 2005, after we received beam time at the newly opened FRMII reactor in Munchen. In this experiment we identified the events of radiative neutron decay and measure its relative intensity B.R.=(3.2±1.6)·$10^{-3}$ with C.L.=99.7% and gamma quanta energy over 35 kev [5, 6]. A year after our discovery of the radiative neutron decay, a NIST experimental group published the results of their experiment on the study of the radiative neutron decay [7] in Nature, with their own value of B.R. = (3.13±0.34)·$10^{-3}$ with C.L.=68% and gamma quanta energy from 15 to 340 kev. This work is dedicated to the comparison of these two experiments.

Our group carried out calculations of the neutron radiative spectrum in the framework of standard electroweak theory about ten years ago [1-4]. The calculated branching ratio for this decay mode as a function of the gamma energy threshold is shown in Figs. 1 and 2. The branching ratio for the energy region investigated here, i.e. over 35 keV, was calculated to be about 2·$10^{-3}$ ( gamma energy threshold ω on Fig. 1 and 2 is equal to 35 keV [3] ).

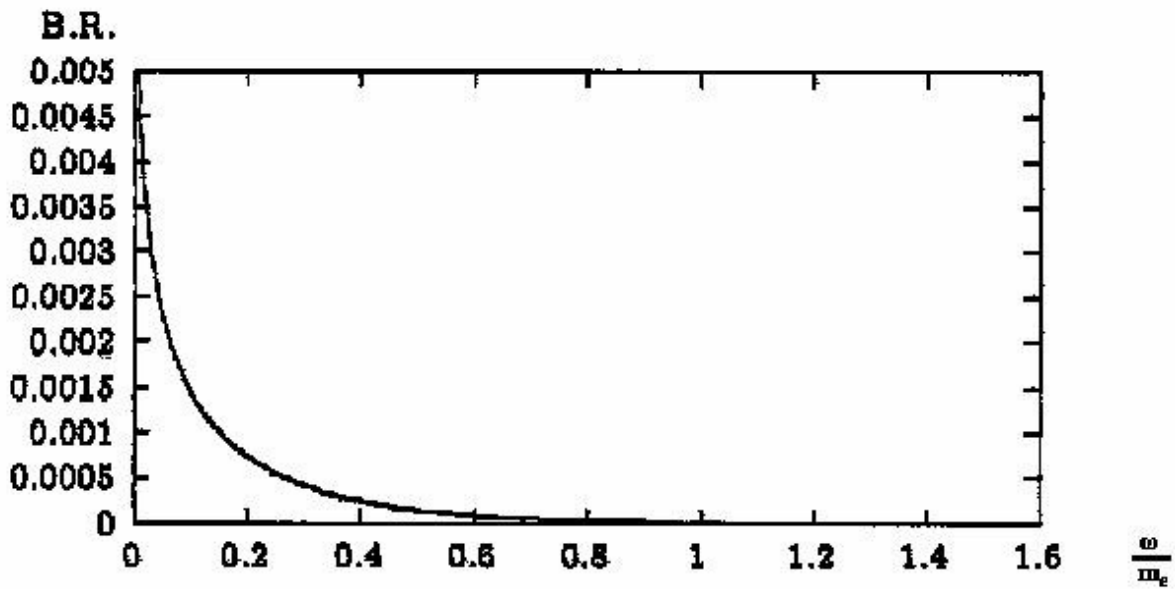

**Fig. 1**. The expected standard model branching ratio for radiative neutron beta decay (summed over all gamma energies larger than the threshold gamma energy ω) as a function of ω (from [1-4]).

Given this rather large branching ratio of about two per thousand, it is in principle not a difficult task to measure it. In practice, however, a rather significant background, mainly caused by external bremsstrahlung emitted by the decay electrons when stopped in the electron detector, has to be overcome. In this experiment this was achieved with the help of a triple coincidence requirement between the electron, the gamma-quantum and the recoil proton. The presence of such a coincidence is used to identify a radiative neutron decay event, whereas an ordinary neutron beta decay is defined by the coincidence of an electron with a recoil proton. The latter coincidence scheme is routinely used to measure the emission asymmetry of electrons in the decay of polarized neutrons [9-10]. Besides the non-correlated background one also has to deal with a correlated background of bremstrahlung gamma-quanta that fully simulates the desired fundamental radiative decay process searched for. This correlated background is caused by bremsstrahlung emission of the electron traveling through the electron detector and is quite significant even when the thickness of this detector is limited to only a few mm. It cannot be eliminated by requiring a triple coincidence of the electron, the photon and the proton. However, calculations [3] show that the radiative emission of a photon in neutron decay is not in the forward direction with respect to the electron emission direction, as in the case of bremsstrahlung, but reaches a maximum intensity at an angle of 35° (Fig. 3). It was this property of radiative neutron decay that led us to use the space solution and register gamma quanta and electrons by different detectors, located at an angle to each other. To achieve this, we constructed a segmented electron-gamma detector with a 35° angle between the sections for electron and gamma detection to reduce this background.

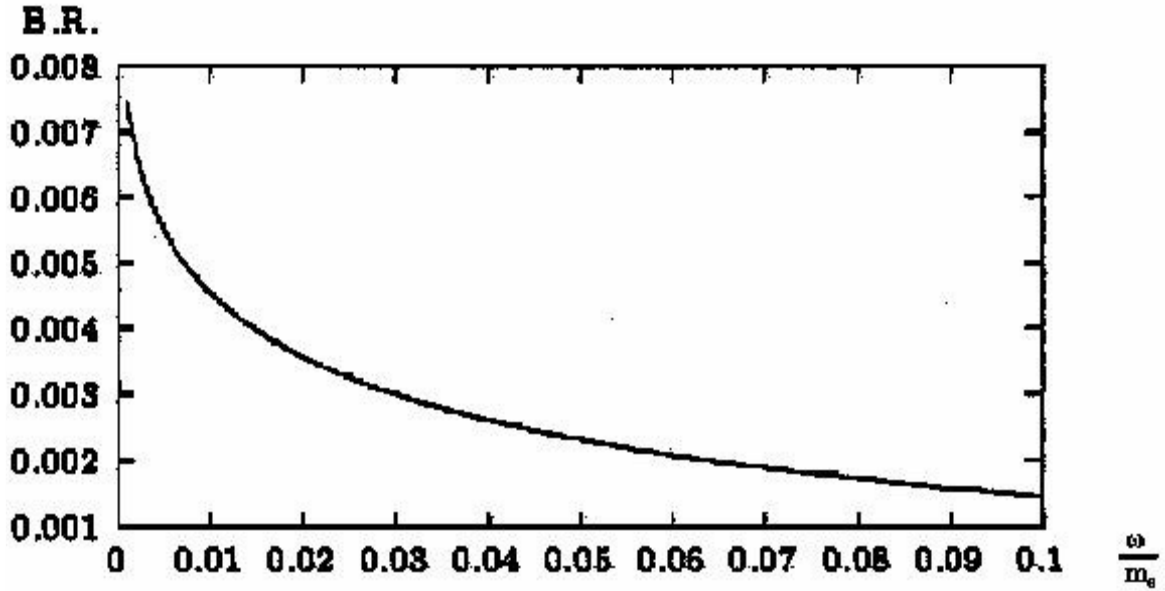

**Fig. 2**. The expected standard model branching ratio for radiative neutron beta decay (summed over all gamma energies larger than the threshold gamma energy ω) as a function of ω, for the low energy part of the spectrum (from [1-4]).

The experimental set-up is shown schematically in fig. 4. The intense cold neutron beam passes through a rather long neutron guide in which is installed a collimation system made of LiF diaphragms, placed at regular distances of 1 meter. The neutrons enter the vacuum chamber (1) through the last diaphragm (9) that is located directly before the decay zone. This zone is observed by three types of detector: the micro channel plate (MCP) proton detector (3), the electron detector (13) consisting of a 7 cm diameter and 3 mm thick plastic scintillator, and six gamma detectors (11) that are located on a ring centered around the electron detector and which consist of photomultiplier tubes each covered with a layer of CsI(Tl) scintillator. The thickness of these 7 cm diameter CsI(Tl) scintillators is 4 mm and has been selected so as to have a 100% detection efficiency for photons. The six gamma detectors (11) surround the electron detector (13) (cf. the lower part of fig. 4) at an angle of 35° and are shielded from it by 6 mm of lead (12). By requiring a coincidence between the electron detector and any of the gamma detectors the bremsstrahlung background can in principle be overcome completely, because bremsstrahlung emission occurs only in the section that registers the electron. In this case, part of the statistics is lost, of course, as can be seen from figure 3. However, the neutron beam intensity of $10^{12}$ n/s in our experimental chamber is sufficiently enough to compensate for that loss and still allows for a good count rate. Recoil protons, formed in the decay zone, pass through a cylindrical time of flight electrode (7) in the direction of the proton detector (3) and are focused onto this detector with the help of spherical focusing electrodes (2). The focusing electrostatic field between the high voltage spherical and cylindrical electrodes (2) and (7) is created by the grids (5) and (6) at one side and by the proton detector grid (4), at ground potential, at the other side. It is important to note that the recoil protons take off isotropic from the decay point. In order not to lose the protons emitted towards the electron detector, an additional grid (10) is added on the other side of the decay volume.

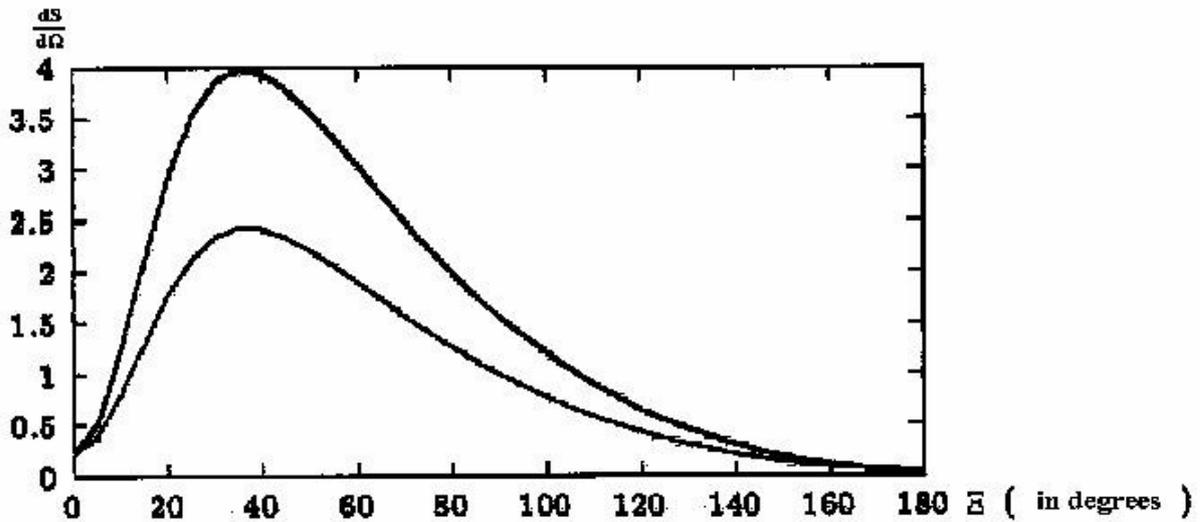

**Fig. 3.** Dependence of the radiative decay spectrum on the angle Ξ between the photon and the electron momenta (upper curve for a threshold gamma energy of 25 keV, lower curve for a threshold gamma energy of 50 keV) (from [3,4]).

The start signal that opens the time windows for all detectors is the signal from the electron, registered in the electron detector (13). For an event to be considered as a radiative neutron decay event there have to be simultaneous signals from the electron detector (13) and one of the gamma detectors (11), followed by a delayed signal from the proton detector (3). It is important to note that in the case of radiative decay, the gamma quantum in our equipment is registered by gamma detectors (11) surrounding the electron detector (13) before the electron is registered by the electron detector. In other words, electron is delayed in comparison to the radiative gamma quantum. In the future namely this fact will allow us to distinguish the peak of radiative gamma quanta in the triple coincidences spectrum. Besides these triple coincidences also double electron-proton coincidences, signaling an ordinary neutron decay event, are monitored.

It is important to note here that the installation of the LiF ceramics diaphragm system in the neutron beam line has significantly reduced the gamma background from the intense cold neutron beam. The background level in the gamma detector amounted to about 2.5 kHz only (at a neutron beam intensity of $10^{10}$ n/s). If the number of the diaphragms in the neutron guide were doubled, the background of the gamma detectors could be further reduced by another order of magnitude, thus becoming comparable to the noise of the photomultiplier tubes. Another important note is that in our last experiment at ILL we succeeded at obtaining a gamma background that was smaller by an order. This could be explained by the fact that on the "Mephisto" beam at FRMII we used a collimation system that was reduced in comparison to the original. Besides, our experiment was the first to be conducted on the intensive cold beam neutron "Mephisto" and as it turned out the axis of this beam went a little higher than the axis of our collimation system, which also contributed to a significant increase in the gamma background. The count rate in the electron detector was just about 100 Hz. It is very likely that most of this count rate is due to electrons from neutron decay since the count rate in this detector almost immediately dropped to zero when the neutron beam was switched off. The value of proton detector background amounted to about 4-6 kHz, which turned out to be very sensitive to the vacuum conditions in the experimental chamber. This background is explained primarily by the presence of a high number of ions in the decay zone, which are captured by the external electric field and are registered by the proton detector (3) along with the recoil proton. ( see Fig. 4 ).

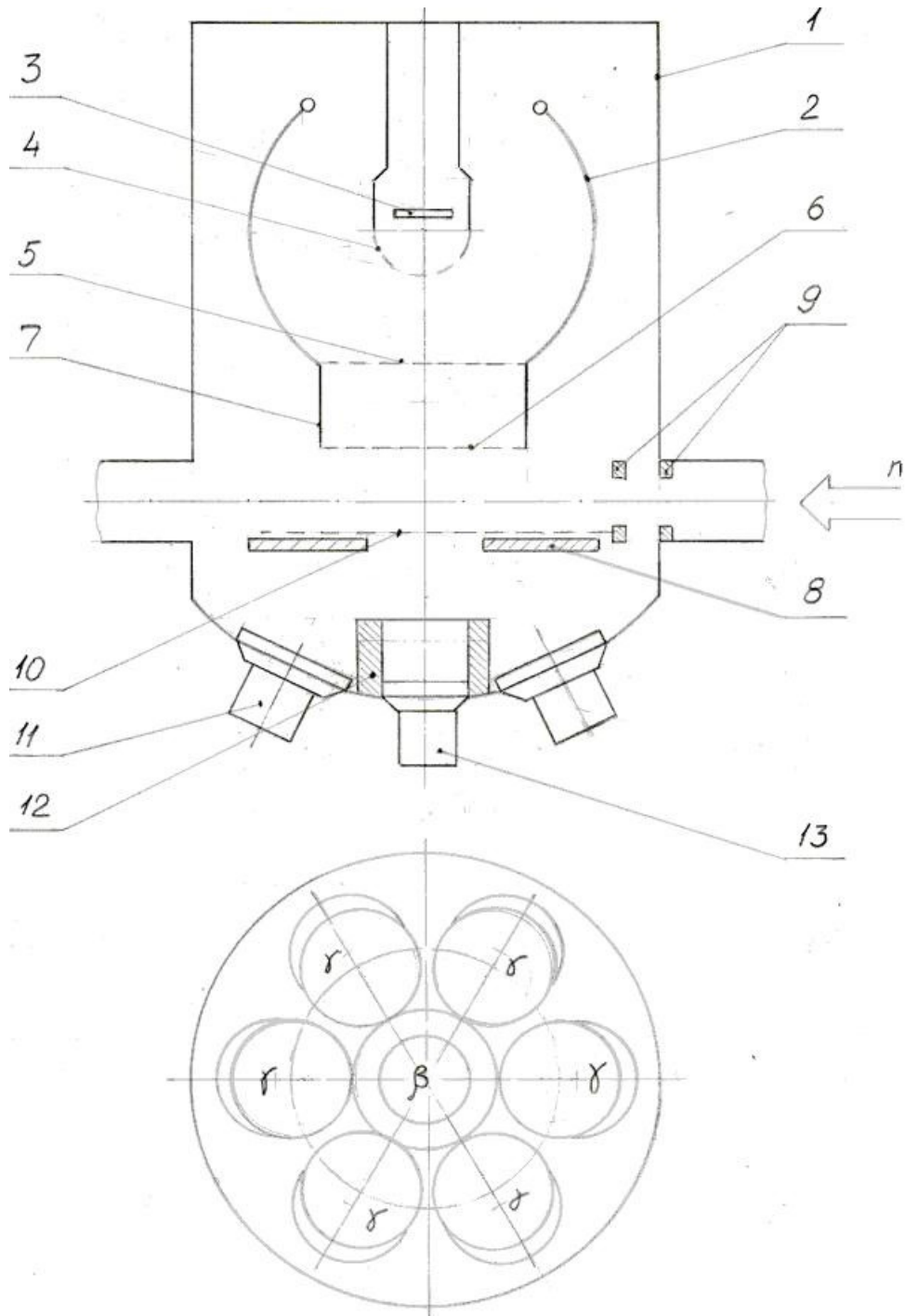

**Fig. 4.** Schematic lay-out of the experimental set-up.
(1) detector vacuum chamber, (2) spherical electrodes to focus the recoil protons on the (at 18-20 kV), (3) proton detector, (4) grid for proton detector (at ground potential), (5) & (6) grids for time of flight electrode, (7) time of flight electrode (at 18-20 kV), (8) plastic collimator (5 mm thick, diameter 70 mm) for beta-electrons, (9) LiF diaphragms, (10) grid to turn the recoil proton backward (at 22-26 kV), (11) six photomultiplier tubes for the CsI(Tl) gamma detectors, (12) lead cup, (13) photomultiplier tube for the plastic scintillator electron detector.

All components of the experimental set-up performed well. Electron-proton coincidences could clearly be observed, while the sectioned electron-gamma detector performed as expected. The results obtained in a vacuum less than $10^{-6}$ mbar are presented in figures 5 and 6. We would like to emphasize that Fig. 5 and 6 of our article present experimental spectra in their raw form, as they were before any preliminary data processing or background subtraction.

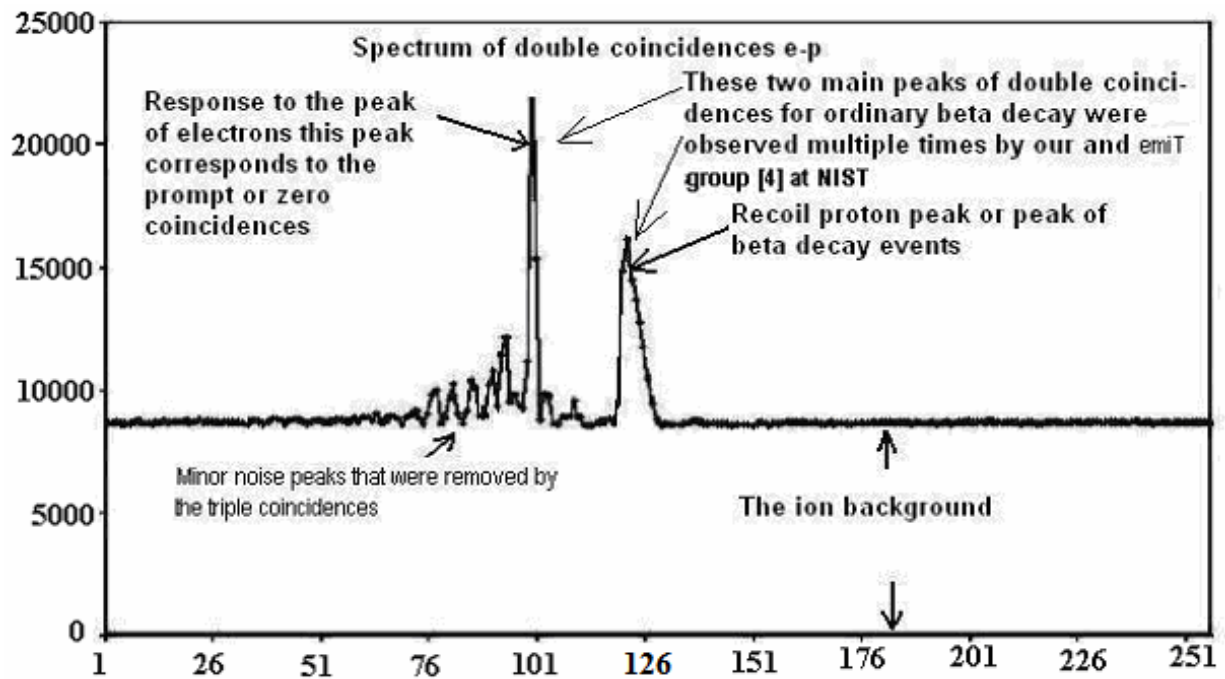

**Fig. 5**. Timing spectrum for e-p coincidences. Each channel corresponds to 25 ns. The peak at channel 99-100 corresponds to the prompt ( or zero ) coincidences. The coincidences between the decay electrons and delayed recoil protons (e-p coincidences) are contained in the large peak centered at channel 120.

Fig. 5 demonstrates the summary statistics on double e-p coincidences (coincidences of electron with delayed proton), and Fig. 6 demonstrates the summary statistics on triple e-p-γ coincidences (coincidences of electron, gamma-quantum, and delayed proton). Fig. 5 clearly shows two major peaks: one peak with a maximum in channels 99-100, which is the peak of zero or prompt coincidences [5, 6]. The position of this peak marks the zero time count, namely the time when the electron detector registered the electron. This peak is not physical in its nature, instead it is a reaction of the detectors and the electronic system to the registration of the beta electron. As we noted above it is namely the pulse from the electron channel that opens the time windows on spectra Fig. 5 for 2.5 μs forward and backwards. The next peak visible on Fig. 5 has a maximum in channel 120 and is the peak of e-p coincidences of electron with delayed proton. An analogous situation was observed in experiments on the measurement of the correlation coefficients by two independent groups at ILL [10] and at NIST [11], and it was also mentioned at [12]. We would especially like to emphasize the correspondence of our spectrum of double coincidences with an analogous spectrum from the result obtained by the emiT group from NIST [11]. On Fig. 6 we present their spectrum and scheme for the registration of the beta electron and the recoil proton. A comparison of our results with the results of the emiT group shows their unquestionable similarity. Moreover, the position of the second proton peak in Fig. 6 (emiT group), like in Fig. 5 (our result), corresponds well to the simple estimate obtained by dividing the length of a proton trajectory to its average speed.

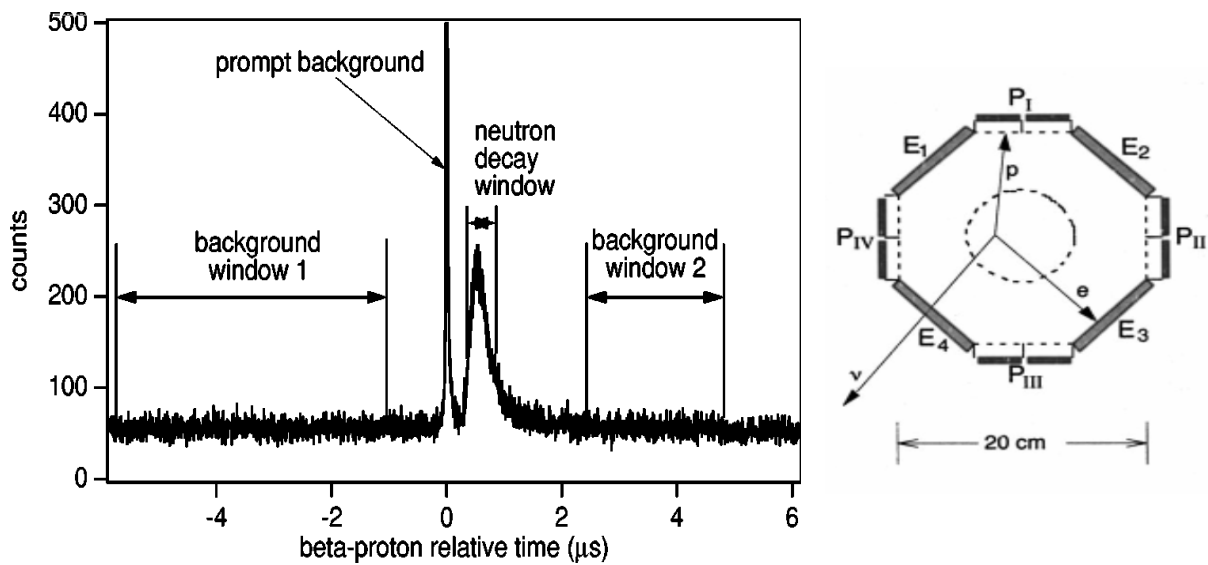

Fig. 6 Spectrum of double electron-proton coincidences obtained by emiT Group [11] with two peaks and ion background value comparable to the neutron decay peak; emiT group scheme for registering beta electron and recoil proton.

Fig. 5 shows that the total number of events in e-p coincidences peak in our experiment equals $N_D=3.75 \cdot 10^5$. This value exceeds the value we obtained in our previous experiment conducted on beam PF1 at ILL by two orders. It was precisely because of the low statistics volume that we could not identify the events of radiative neutron decay in that experiment and defined only the upper B.R. limit [8]. It is very important to note that the peak of double coincidences between electron and the delayed proton is observed against a non-homogenous background: besides the homogenous ionic background, which has a value comparable to the value of the e-p coincidences peak, there is an obvious peak in channels 99-100. In essence, this peak is a response peak to the time spectrum of electron registration, which contains just one peak in channels 99-100, signifying the time when the electron detector registered the electron. In the future we will see that the radiative peak of triple coincidences appears against a non-homogenous background with not one, but two response peaks.

The remaining peaks on Fig. 5 are small, with just seven peaks distinct from the statistical fluctuations. These occurred because of the noise in the electric circuits of the FRMII neutron guide hall. There are no other physics-related reasons for their occurrence. These peaks appeared and disappeared depending on the time of day, reaching their maxima during the work day and disappearing over the weekends. Such behavior was observed throughout the experiment as we collected statistics. Since the nature of these seven small peaks is in no way related to radiative and ordinary decay, we did not emphasize them in our article.

Analysing the double coincidences spectra obtained by our and the emiT groups (both of which present two main peaks) shows that in the spectrum of triple coincidences we should observe not two but three peaks. Namely, along with the sought after radiative peak, the triple e-p-gamma coincidences spectrum should show two response peaks to the registration of beta-electrons and the registration of protons. Fig. 7 of triple coincidences clearly shows three peaks, and the leftmost peak with the maximum in channel 103 is connected to the peak of the radiative gamma-quanta in question, as this gamma-quantum is registered by the gamma detectors in our equipment before the electron. Comparing Fig. 5 and 7, it becomes clear that if we ignore the first leftmost peak with the maximum in channel 103 in Fig. 7, the spectrum of double e-p coincidences will resemble the spectrum of triple e-p-γ coincidences. The peak with the maximum in channel 106 on Fig. 6 is connected to the left peak of false coincidences on Fig. 5, and the peak with the maximum in channel 120 on Fig. 7 is connected to the right peak of e-p coincidences on Fig. 5. The emerging picture becomes obvious when one uses a standard procedure, introducing a response function for gamma channel $R_\gamma(t,t')$, which is also necessary for calculating the number of triple radiative coincidences $N_T$ in radiative peak. Using the method of response function, one can confidently define our double-humped background: the narrow peak with the maximum in channel 106 on Fig. 7 is the response to the narrow peak of zero coincidences in channels 99-100 on Fig. 5, and the second peak in this double-humped background on Fig. 7 is the response to the peak in channels 117-127 on Fig. 5.

Of course, the width of these response peaks is greater than the width of the two original peaks on Fig. 5: the width of the narrow peak increases more than width of wide peak, the distance between the peaks themselves diminishes, and the narrow peak moves to the right towards the wider peak more than this wide peak moves to the left. Such behavior is described using the standard method of response function, and there is nothing unusual about it. Indeed, for real detectors and electronics the response function is always not local, which explains all these deformations. In our case this non-local behavior is connected to the fact that the rise time of the gamma signal from gamma detectors is on average 150-200 ns (which equals to 6-8 time channels on Fig. 5 and 7), and the rise time of the electron signal from the electron detector equals to 10 ns. This difference explains, in particular, the shift of the peaks on Fig. 7: the peak with maximum in channel 106, which is the response to the peak of zero coincidences, is shifted to the right (the side of delay) in comparison to the peak of zero coincidences in channel 99-100 on Fig. 5.

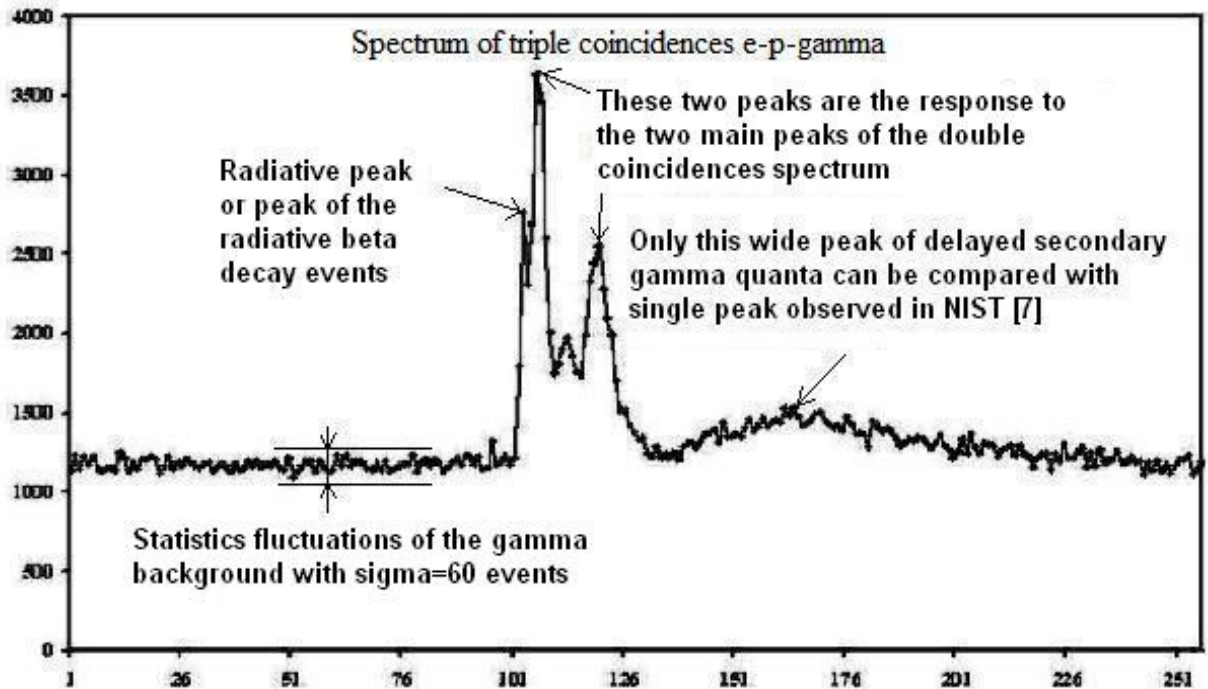

**Fig. 7.** Timing spectrum for triple e-p-g coincidences. Each channel corresponds to 25 ns. In this spectrum, three main peaks in channels 103, 106 and 120 can be distinguished. The leftmost peak in 103 channel among these three main peaks is connected with the peak of radiative decay events.

Here it is also appropriate to remark on another peculiarity observed throughout the experiment and which emphasizes the physics-related nature of the peak in channel 103 on Fig. 7. As noted above, the noise peaks on Fig. 5 were not stable, and neither were the small peaks in channels 96 and 116 on Fig. 7. Besides, if the small noise peaks in the spectrum of double coincidences disappeared, the small peaks in channels 96 and 116 in the spectrum of triple coincidences disappeared as well. The radiative peak in channel 103 was stable throughout, it never "migrated" to a different channel and its growth was stable, regularly collecting the same number of events during the same stretch of time!

As for the wide, almost indistinguishable peak in channel 165 on Fig. 7, its influence on radiative peak in channel 103 is negligible. Its nature is in no way related to the researched phenomenon, so we do not discuss it in our article. This peak is created by the radioactive gamma quanta delayed on average by 1.25 µs and emitted by the radioactive medium within our experimental equipment. The medium is activated by registered beta-electrons. This event of artificial, induced radioactivity has been known for over 100 years and does not have anything in common with the new event of radiative neutron decay which is the subject of current research. As we will demonstrate below, only this 1 microsecond peak and delayed from the registration time by about the same time can be compared to the peak observed by the authors of paper [7] at NIST. Thus, the authors of this experiment observed not the events of radiative decay but rather the event of artificial radioactivity, already well known in the time of Joliot-Curie.

After analyzing the spectra with the help of the non-local response function we finalize the average value for the number of radiative neutron decays $N_T=360$ with a statistics fluctuation of 60 events. B.R. can be expressed through ratio $N_T$ to $N_D$ as BR = k $(N_T / N_D)$, where coefficient k=3.3 is geometrical factor what we can calculate by using not isotropic emission of radiative gamma-quanta Fig. 3. With the number of observed double e-p coincidences $N_D = 3.75 \cdot 10^5$ and triple e-p-$\gamma$ coincidences $N_T = 360$, one can deduce the value for radiative decay branching ratio of $(3.2 \pm 1.6) \cdot 10^{-3}$ (99.7 % C.L.) with the threshold gamma energy $\omega=35$ keV. The average B.R. value we obtained deviates from the standard model, but because of the presence of a significant error (50%) we cannot make any definite conclusions. The measurements must be made with greater precision. According to our estimates, we will be able to make more definite conclusions about deviation from the standard electroweak theory with experimental error less than 10%.

Unfortunately, the other report [7] published a year after our article and dedicated to the experimental study of the radiative neutron decay mode with which we are comparing our results publishes neither initial spectra with raw experimental data on double coincidences of β-electron with recoil proton, nor data on triple coincidences of the electron-proton pair with the radiative gamma-quantum. In and of itself, this undermines confidence in this report. The only initial experimental data published in the report are the shape of the pulses from the combined electron-proton detector and the gamma-quanta detector (see Fig. 8 from [7]).

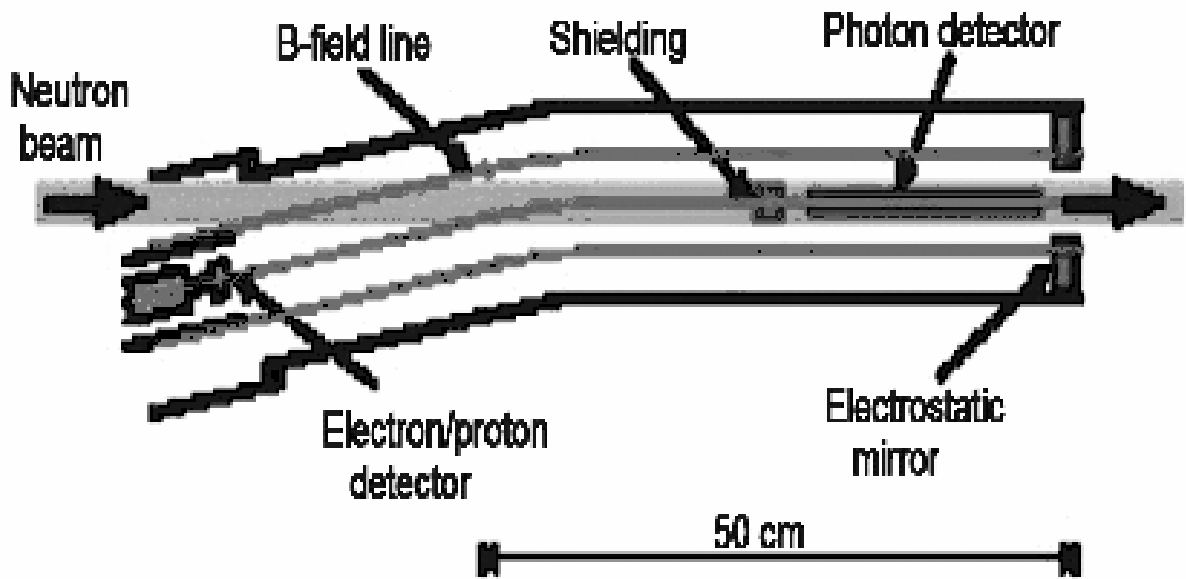

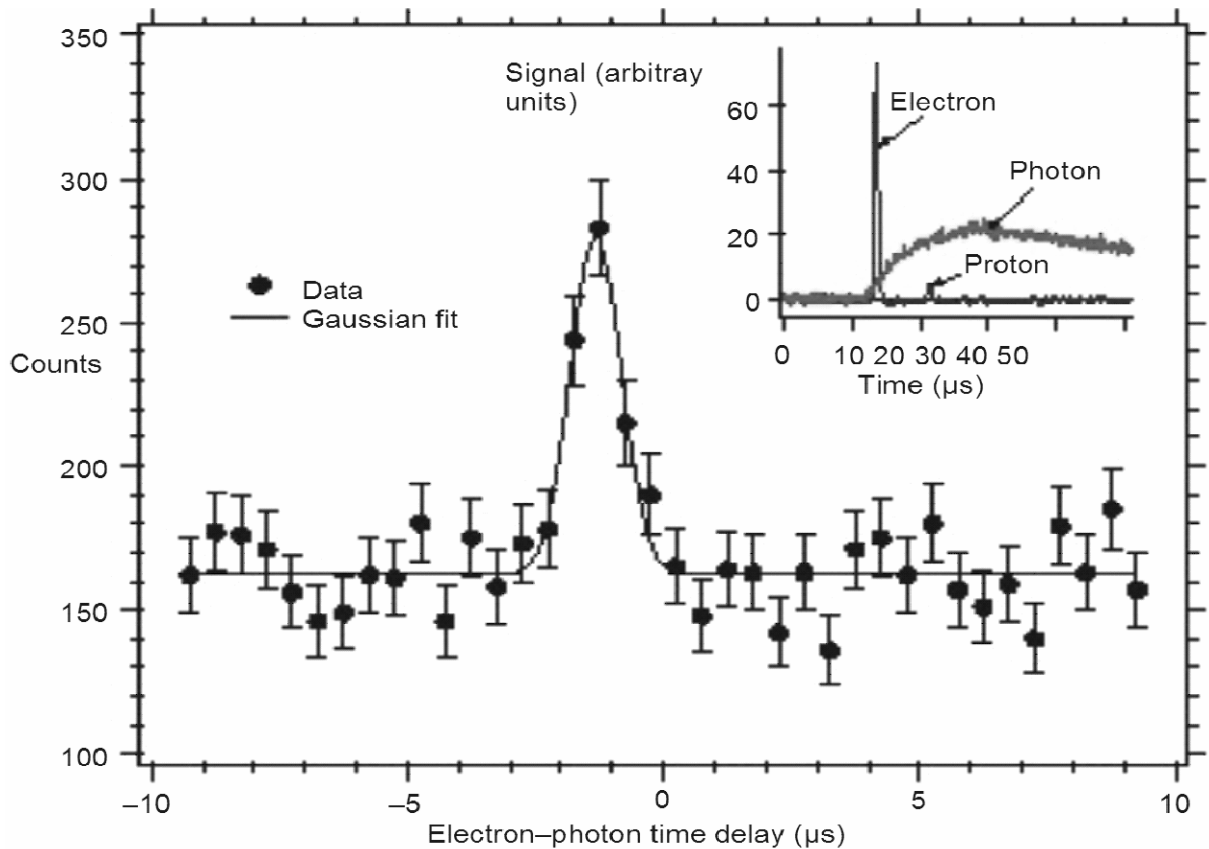

Fig. 8 Equipment diagram and the only peak of "electron-photon" coincidences, sihfted to the left of 0 – the time of beta-electron registration – by 1.25 microseconds, published in [7].

The difference between the NIST experiment and our experiment becomes immediately apparent. First and foremost, it is the time scale: in our spectra, the scale is measured in nanoseconds, while in the other experiment the scale is in microseconds. Besides, we used three types of detectors, each of which registered its own particle: one detector for the electrons, one for the protons, and six identical detectors for the radiative gamma-quanta (see Fig. 4). The duration of the front pulse from the electron and proton detectors is 10 nanoseconds in our experiment and 100 times greater than that in the NIST experiment, in the order of 1 μs. The rise time of gamma signal from our gamma-detectors is on average 150 ns, and from avalanche diode on the NIST equipment greater than 10 μs, besides that the diode pulse arrives with significant noise, which makes the thickness of the front pulse line equal to more than 0.5 μs (see the photon line on Fig. 8 from [7]). All of this leads to our factual time resolution being two orders better than the resolution achieved in the NIST experiment. However, as the two experiments used equipment which was practically the same in size and smaller than 1 meter, the choice of the time scale is a matter of principle. Given this geometry, it is impossible to get microsecond signal delays from all of the registered charged particles, i.e. electrons and protons. In this light, it is surprising that the peak identified by the authors of the NIST report [7] as the peak of radiative gamma-quanta, is shifted by 1.25 microseconds to the left. The expectation that magnetic fields of several tesla in magnitude delay all electrons and protons, are absolutely ungrounded.

Indeed, the electron and proton velocities depend only on the kinetic energy of these particles. Fig. 9 shows the dependence of the ratio of electron velocity v to the speed of light c on the ratio of electron kinetic energy $T_e$ to the electron mass energy $m_e c^2$. It is evident that the speed of β-electron, the kinetic energy of which is measured in hundreds of keV, is comparable with the speed of light and in one microsecond should travel hundreds of meters, and not one meter. The length of trajectory in magnetic fields also does not depend on the magnitude of the magnetic field, and rather depends only on the direction of electron emission in relation to the direction of the magnetic field. Fig. 10 shows the dependence of the ratio of trajectory length l to the distance from the point of decay to detector L on angle θ between the direction of electron emission and the direction of the magnetic field. This dependence is rather smooth and also does not depend on the magnitude of the magnetic field. As demonstrated by this graph, the ratio is l/L < 2 for electrons emitted at angle θ less than $60^0$ and l/L < 10 for θ<$85^0$ . As β-electron emission is isotropic, it means that practically all electrons travel from the point of decay to the detector in tens of nanoseconds, but not in microseconds.

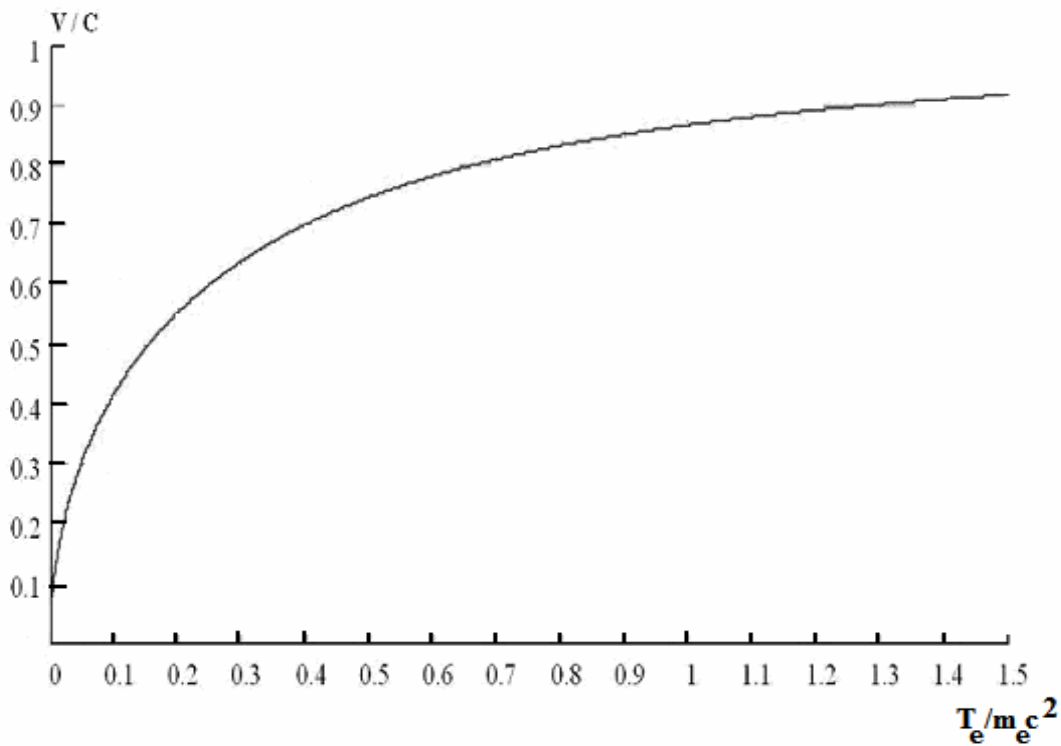

**Fig. 9.** Relationship between the ratio of electron speed v to the speed of light c and the ratio of kinetic electron energy Te to the energy of its mass $m_e c^2$.

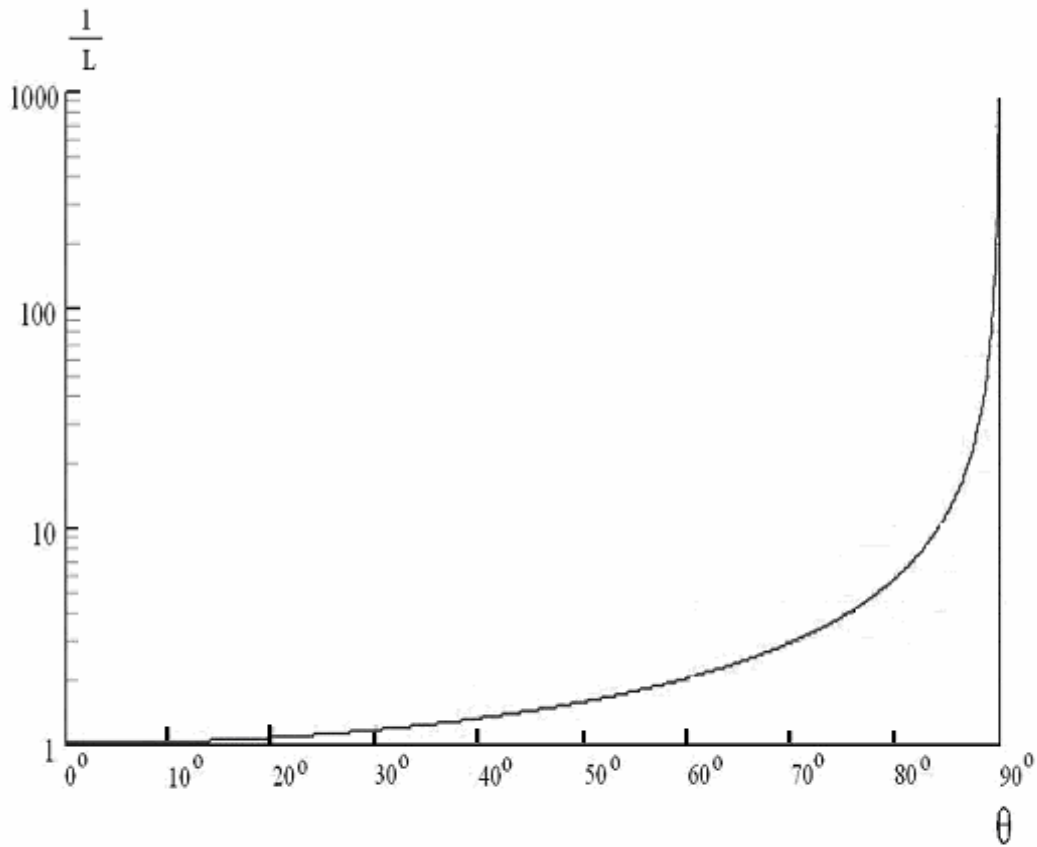

**Fig. 10.** Relationship between the ratio of the trajectory length l of charged particles moving in the magnetic field to distance L between the point of decay and the detector and the angle between the velocity of this particle and the direction of the magnetic field θ.

When we discuss the delay of beta-electrons in relation to radiative gamma-quanta, we would like to emphasize the following point in methodology. We wrote about this directly in our article published in JETP Letters [5], when discussing the reasons for the non-locality of the response function. This aspect concerns the method of determining the location of the radiative peak in relation to the time of electron registration. The beginning of the pulse from the radiative gamma-quantum obviously predates the beginning of the pulse from the electron detector. But, as we just explained, this difference is extremely small and is measured in nanoseconds, and not microseconds. However, noises do not allow to exactly determine the location of the beginning for the gamma-quantum signal, so we determined the location of the radiative peak using the true-constant-fraction ( TCF ) method in our coincidences scheme, using a constant fraction discriminator ( CFD ) [13]. As the fronts of the gamma detector signals are flatter than the fronts of the electron detector signals, the time-pickoff signal from the gamma quanta appeared later in relation to the signal of the electron time-pickoff signal, so the location of the radiative peak should be to the right of the electron registration time. In other words, the radiative peak in our coincidences scheme should be delayed in relation to the electron registration time by the electron detector. In our case (see Fig. 7), the radiative peak is located to the right of the 100th channel by a value less than the rise time of the gamma-detector signal and is positioned with precision of 25 ns. The response to the peak of momentary coincidences in the spectrum of triple coincidences ( the peak with the maximum in channel 106 on Fig. 7 ) is also delayed and is shifted to the right of the 100th channel. At the same time, in the spectrum of triple coincidences on Fig. 7 the radiative peak is located, as it should be since the radiative quantum arrives at the gamma detector before all other particles, on the left slope of the response peak, and not on the right. Thus, we obtained the location of our radiative peak exactly where we expected it. As it has already been noted above, the movement of the response peaks and their widths are accounted for in the analysis of experimental data by the introduction of the non-local response function.

Now let us consider the other coincidence pairs, namely the coincidences of electrons and recoil protons. Fig. 11 presents t the dependence of the ratio of proton velocity v to the speed of light c on the ratio of proton kinetic energy $T_p$ to the proton mass energy $m_p c^2$ As the original kinetic energy of the recoil protons in the beta-decay is extremely low, it is determined primarily by the potential of the electron-proton detector (in our case, this is the potential of focusing electrodes). However, even in the case where this potential is equal to just 35 kev, proton velocity is still significant and is slightly less than 1% of the light velocity **c**. In our case, when proton kinetic energy was 25 keV, the final velocity was 0.006 c, and its average velocity along the trajectory was 0.003 **c**. On Fig. 5, which shows the spectrum of the electron-proton coincidences, the peak of these coincidences is located in the 120th channel, which corresponds to proton delay of 500 ns on average or the distance between the point of decay and the proton detector in 40-50 cm. The estimate obtained is quite good and coincides with the real distance between the proton detector and the axis of neutron beam in our equipment with precision of ten-twenty per cent. The same can be said about the delay of protons vs. electrons, measured by the emiT group from NIST (see the distance between the first and the second peak on their spectrum, Fig. 6). There, the delay also corresponds well to the estimate received by simply dividing the length of the proton trajectory by its average speed within the experimental equipment.

The presence of a strong magnetic field cannot radically change the obtained value of the recoil proton delay and change it from hundreds of nanoseconds to ten microseconds.

As the initial energy of the recoil proton is very small in comparison to electric field potential, angle θ (see Fig. 3) is now defined as the angle between the direction of the electric field and the magnetic field. As follows from the geometry of the NIST experiment equipment, the electric field is directed along the magnetic field and angle θ should be no more than a few degrees. On the other hand, as follows from our spectrum of double e-p coincidences Fig. 5, the peak of electron-proton coincidences is observed with a rather significant ionic background, the value of which is equal to the value of the electron-proton field itself. In other words, the process of ion formation is comparable in its magnitude to the process of the recoil proton formation in the neutron beta-decay. An analogous, although slightly better, situation can be observed with the ionic background obtained by the emiT group (see Fig. 6). It means that these ions may be captured by the strong magnetic field in our opponents' equipment and produce pulses delayed by ten microseconds. From this it follows that the signal from recoil protons on Fig. 11 is at the base of the electron signal the duration of which is over 1 microsecond, and so simply cannot be registered by the combined electron-proton detector in the NIST experiment equipment. The signal shown on Fig. 11 as a proton signal delayed by 10 microseconds, and is the signal from an ion captured by the magnetic and electric fields. So, the use of the strong magnetic fields prevents reliable identification of ordinary neutron decay events, much less radiative events, in the NIST experiment. Namely for this reason the authors of NIST experiment did not publish their spectrum of double e-p coincidences in Nature - they simply did not observe the peak of neutron decay as it was seen in our spectrum on Fig. 5 and emiT group spectrum on Fig. 6.

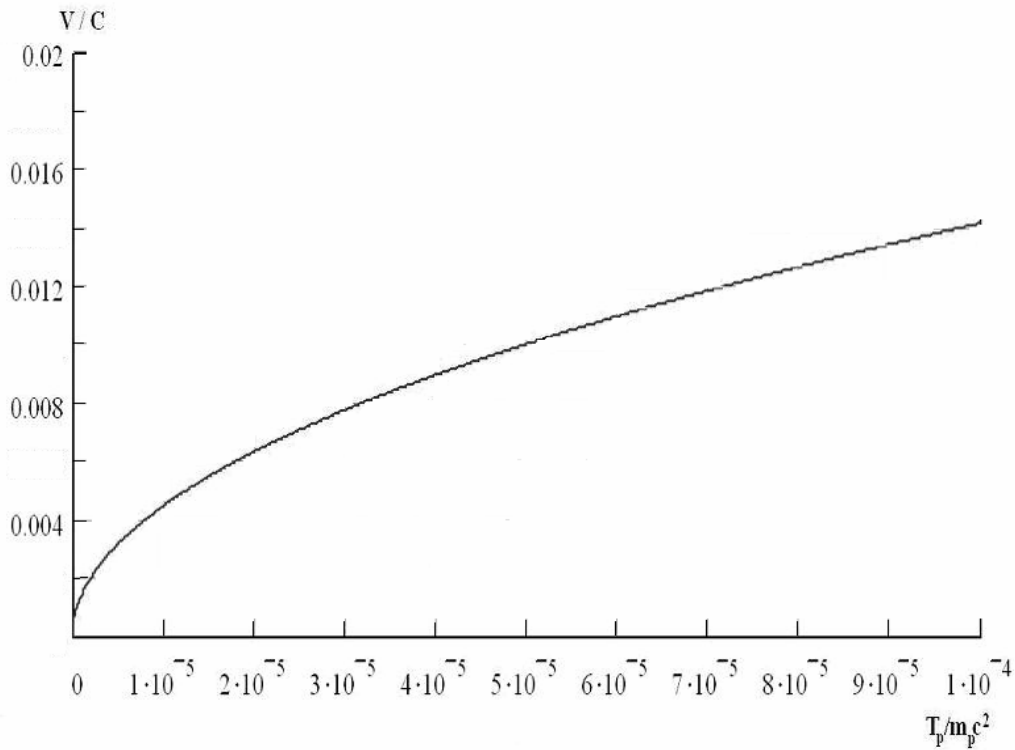

**Fig. 10.** Relationship between the ratio of proton velocity v to speed of light c and the ratio of kinetic proton energy $T_p$ to its mass $m_p c^2$.

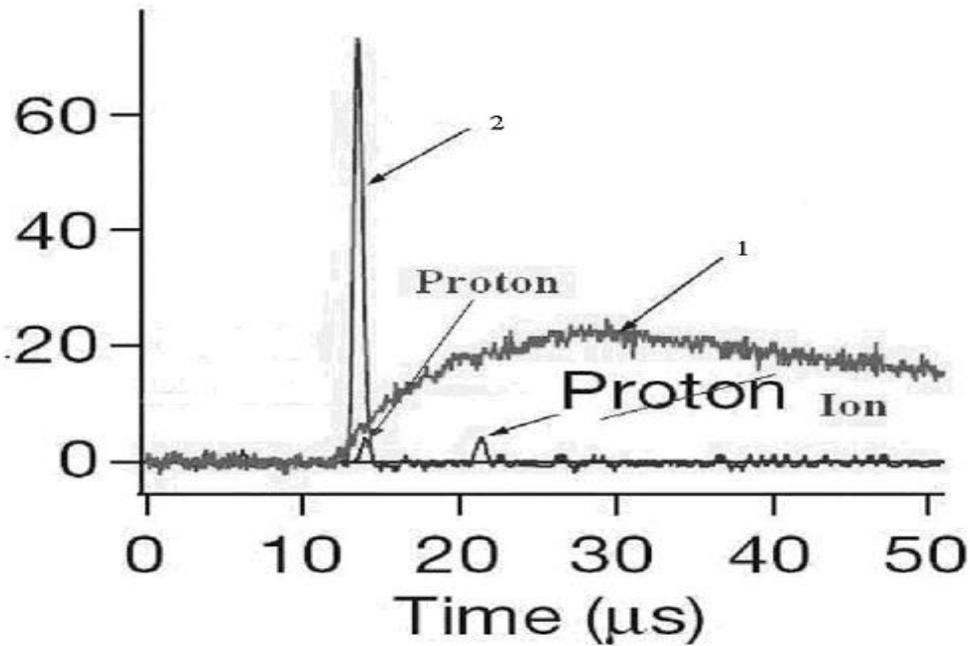

**Fig. 11**. The signal from the decay proton has to be delayed by less than one microsecond, which is why it is located at the base of the electron pulse ( see line number 2 ) and so cannot be registered by the combined electron-proton detector. The pulses that are delayed by longer than 1 microsecond are pulses not from decay protons, as it was indicated in ref. [7], but rather from ions, formed in the decay zone. The line number 1 shows the shape of pulses from the gamma detector

Finally, it is necessary to note that when using strong magnetic fields it is impossible to determine the average angle between electron moment direction and the direction of radiative gamma-quantum momenta. As shown in the calculated spectrum on Fig. 3, the intensity of radiative gamma-quanta emission strongly depends on this angle, which makes comparison between theory and experimental data, obtained in magnetic fields, very problematic. In our equipment, on the other hand, the average angle between the electron and the gamma-quanta momenta is fixed in the region of maximum radiative gamma quanta emission intensity and so determines the geometric factor.

In conclusion, we would like to emphasize two main points concerning the experiment we conducted.

First, the results from the first experiment aiming to observe the as yet undiscovered radiative decay mode of the free neutron are reported. Although the experiment could not be performed under ideal conditions, the data collected still allowed one to deduce the B.R. = $(3.2\pm1.6) \cdot 10^{-3}$ (99.7 % C.L.) for the branching ratio of radiative neutron decay in the gamma energy region greater than 35 keV. This value is in agreement with the theoretical prediction based on the standard model of weak interactions.

Secondly, the average B.R. value we obtained deviates from the standard model, but because of the presence of a significant error (50%) we cannot make any definite conclusions. Taking into account the fact that the experimental conditions can still be significantly optimized, an e-p coincidence count rate of 5-10 events per second is within reach. Together with the standard model prediction for the branching ratio of this decay mode, this would correspond to a triple e-p-γ coincidence rate of several events per 100 seconds. This can easily be observed with the current experimental set-up, which is now being optimized with a view to performing such an experiment. The aim of that experiment will then not only be to establish the existence of radiative neutron beta decay, but also to study B.R. in more detail.

Our first priority for the new experiment will be to suppress the response peaks on triple coincidences spectrum. With this aim we created a multichannel pulse generator which can fully simulate both the load rate of each channel of our electron system and the form of the pulses themselves. We tested our system using this generator and obtained response peaks analogous to what we observed during the real experiment in the spectrum of triple coincidences (see Fig. 5 and 7). Consequently we were able to create a new electron system that allows to substantially suppress the influence of these response peaks to the radiative peak. According to our estimates, this will allow us to measure B.R. with precision of several percentage points in a new experiment. This precision will allow us to either confirm or reject the deviation of the average B.R. value from the standard model, which we observed in our last experiment.

As concerns the comparison of our experimental results with others we can make the following two main conclusions:

The main characteristics (position and width of two peaks, levels of ion background before and after electron registration, etc.) of our spectrum of double electron-proton coincidences identifying the events of ordinary neutron decay fully coincide with an analogous spectrum published by emiT group in [11] and with estimations. It is very satisfying for us to observe the similarity of our results.

Unfortunately we cannot say same for another experiment published in Nature [7]. First of all it is necessary to mention that authors of this article have published neither the spectra of double e-p coincidences nor the spectra of the triple e-p-gamma coincidences, necessary for the measurement of the BR. As we explained above, when using strong magnetic fields makes it impossible to obtain these spectra, as they make it impossible to identify not only radiative events but also the events of ordinary neutron decay. The only peak published in nature is the peak of "electron-photon" double coincides, which the authors use to identify events of radiative decay, is extremely surprising not only because the coincidence of electron with gamma-quantum does not fall under the definition of BR but also because of the characteristics of this peak. Particularly vexing is the authors' unsubstantiated assertion that they observe their only wide peak of gamma quanta before the registration of beta-electrons. Both the position and the width of this peak are located in sharp contradiction to both the elementary estimates and the results of our experiment. In the course of our entire experiment we did not observe such a wide peak in the triple coincidences spectrum, located before the arrival of electrons at a huge distance of 1.25 μs. However, it is possible to reconcile our spectra of triple coincidences with the one isolated peak observed at NIST if we assume that at NIST, the gamma-quanta were registered after the beta electrons. Only in this case does the NIST peak almost completely coincide with the peak we observed in the spectra of triple coincidences with the maximum in channel 163, both in terms of the huge delay of 1.25 μs and in terms of its huge width. This peak is created by the delayed secondary radioactive gamma-quanta, arising from the activation by beta electrons of the media inside experimental chamber. Thus, the authors of this research did not observe radiative neutron decay but rather researched the well known occurrence of artificial, induced radioactivity.

Despite the recent disagreements [14], which we consider to be subjective in nature [15], we acknowledge the contribution of our Western colleagues Profs. N. Severijns, O. Zimmer and Drs. H.-F. Wirth, D. Rich to our experiment conducted in 2005. Here it is important to note that the authors of the article published in Nature [7] consciously misled first our Western colleagues and then the physics community at large by insisting that their only wide peak is removed by 1.25 microseconds to the left from the time of electron registration, when in reality this peak was formed by delayed gamma-quanta, emitted by the activated medium inside the experimental equipment, and corresponds to our wide peak with the maximum in channel 165 (refer to Fig. 7) [14, 15].. The authors would like to thank Profs. D. Dubbers and Drs. T. Soldner, G. Petzoldt and S. Mironov for valuable remarks and discussions. We are also grateful to the administration of the FRMII, especially Profs. K. Schreckenbach and W. Petry for organizing our work. We would especially like to thank RRC President Academician E.P. Velikhov and Prof. V.P. Martem'yanov for their support, without which we would not have been able to conduct this experiment. Financial support for this work was obtained from INTAS (Project N 1-A –00115; Open 2000), the RFBR (Project N 07-02-00517). We would especially like to note the early financial support from the INTAS fund. Without this support neither the 2002 ILL experiment, nor the 2005 experiment at TUM which we discuss in this paper would have been possible.

**References**
[1] Radiative neutron beta-decay and its possible experimental realization
Gaponov Yu.V., Khafizov R.U. Phys. Lett. B 379 (1996) 7-12
[2] Radiative neutron beta-decay and experimental neutron anomaly problem. By Yu.V. Gaponov, R.U. Khafizov . Weak and electromagnetic interactions in nuclei (WEIN '95): proceedings. Edited by H. Ejiri, T. Kishimoto, T. Sato. River Edge, NJ, World Scientific, 1995. 745p.
[3] About the possibility of conducting an experiment on radiative neutron beta decay
R.U. Khafizov, N. Severijns Proceedings of VIII International Seminar on Interaction of Neutrons with Nuclei (ISINN-8) Dubna, May 17-20, 2000 (E3-2000-192), 185-195
[4] Angular distribution of radiative gamma quanta in radiative beta decay of neutron.
Khafizov R.U. Physics of Particles and Nuclei, Letters 108 (2001) 45-53
[5] R.U. Khafizov, N. Severijns, O. Zimmer et al. JETP Letters 83 (2006) p. 5
[6] Discovery of the neutron radiative decay R.U. Khafizov, N. Severijns et al. Proceedings of XIV International Seminar on Interaction of Neutrons with Nuclei (ISINN-14) Dubna, May 24-25, 2006
[7] J. Nico, et al. Observation of the radiative decay mode of the free neutron. Nature, v. 444 (2006) p. 1059-1062
[8] M. Beck, J. Byrne, R.U. Khafizov, et al., JETP Letters 76, 2002, p. 332
[9] B. G. Yerozolimsky, Yu. A. Mostovoi, V P. Fedunin, *et al.*, Yad. Fiz. 28, 98 (1978) [Sov. J. Nucl. Phys. **28**, 48 (1978)].
[10] I.A. Kuznetsov, A. P. Serebrov, I. V. Stepanenko, et al., Phys. Rev. Lett., 75 (1995) 794.
[11] L.J. Lising, S.R. Hwang, J.M. Adams, et al., Phys. Rev. C. v.6, 2000, p. 055501
[12] J.Byrne, R.U. Khafizov, Yu.A. Mostovoi, et al., J. Res.Natl. Inst. Stand. Technol. 110, p.415, (2005).
[13] T.J. Paulus, IEEE Transactions on Nuclear Science, v.NS-32, no.3, June, p.1242, 1985.
[14] N. Severijns, et al., e-print arXiv:nucl-ex/0607023; J. Nico, private communication.
[15] R.U. Khafizov, V.A. Solovei, e-print arXiv:nucl-ex/0608038.18